\documentclass[iop, twocolumns, amssymb,notitlepage]{revtex4-1}
\pdfoutput=1 

\usepackage{graphicx} 
\usepackage{dcolumn}  
\usepackage{bm}       
\usepackage{amsmath}
\usepackage{amssymb}
\usepackage{hyperref}

\usepackage{graphicx}

\usepackage[utf8]{inputenc}
\usepackage[english]{babel}
\usepackage{enumerate}

\usepackage{bm}

\DeclareRobustCommand{\vect}[1]{
  \ifcat#1\relax
    \boldsymbol{#1}
  \else
    \mathbf{#1}
  \fi}
  
 \newcommand{\indice}{n}
 \newcommand{\hh}{B}

\begin{document}

\title{Negative-temperature Fourier transport in one-dimensional systems}
\author{Marco Baldovin}
\affiliation{Department of Physics, Universit\`a ``Sapienza'', Roma  
Piazzale A. Moro 5, I-00185 Italy}
\author{Stefano Iubini}
\affiliation{Consiglio Nazionale delle Ricerche, Istituto dei Sistemi Complessi, via Madonna del Piano 10, I-50019 Sesto Fiorentino, Italy}
\affiliation{Istituto Nazionale di Fisica Nucleare, Sezione di Firenze, 
via G. Sansone 1 I-50019 Sesto Fiorentino, Italy}

\date{\today}

\begin{abstract}
We investigate nonequilibrium steady states in a class of one-dimensional diffusive systems that can attain negative absolute temperatures. The cases of a paramagnetic spin system, a Hamiltonian rotator chain and a one-dimensional discrete linear Schr\"odinger equation are considered. 
 Suitable models of reservoirs are implemented to impose given, possibly negative, temperatures
at the chain ends. We show that a phenomenological description in terms
of a Fourier law can consistently describe unusual transport regimes where the temperature profiles are entirely or partially in the negative-temperature region. Negative-temperature Fourier  transport  is observed both for deterministic and stochastic dynamics and it can be generalized to coupled transport when two or more thermodynamic currents flow through the system.

\end{abstract}

\maketitle 

\section{Introduction}
\label{sec:Intro}

The characterization of steady states of open classical and quantum systems 
is a central problem in physics, with many implications both for theoretical studies and for applications.
In the context of  nonequilibrium statistical mechanics, the study of  transport problems in  simple low-dimensional lattices is a field which has been deeply investigated in the last 
decades. Relevant examples include the discovery 
of nonequilibrium phase transitions~\cite{derrida98, tauber17}, anomalous transport~\cite{lepri03} or mechanisms of
dynamical arrest of transport~\cite{pino16}.

In the typical nonequilibrium setup, the system is steadily driven out of equilibrium by external mechanical or
thermodynamical forces. When the departure 	 from equilibrium is sufficiently small,
the usual assumption is that  macroscopic currents are linearly related to the applied forces.
 The Fourier law is certainly one of the most celebrated examples of such phenomenological relations, where the heat current is proportional to the opposite of the temperature gradient, the proportionality constant being the 
thermal conductivity. 
Despite our understanding of the microscopic origin of  Fourier law  is still far from being complete, 
significant achievements have been made in the last years to explain its emergence or breakdown
in low-dimensional systems~\cite{lepri16}. 

In this paper, we  focus on the problem of heat conduction for a peculiar class of models capable to
attain negative absolute temperatures. Thermodynamics and statistical mechanics at negative absolute temperature have been quite a debated problem in the last decades. Since the pioneering studies by Onsager~\cite{onsager49}
and Ramsey~\cite{ramsey56}, many theoretical and experimental results have been found. Recent examples include
spin systems~\cite{oja97}, vortices~\cite{yatsuyanagi05,pakter18} and cold atoms~\cite{braun13}.
Although a general theoretical framework is reasonably well established for what concerns
equilibrium thermodynamics and relaxation to equilibrium, the problem of steady transport
in the presence of negative temperature states has received much less attention.
In~\cite{iubini17entropy} it was shown that a one-dimensional Discrete Nonlinear Schr\"odinger Equation in
steady out-of equilibrium conditions can spontaneously enter the negative-temperatures region
even when the reservoirs impose positive temperatures at its boundaries. In this example, negative temperature states
are obtained in the bulk of the system and are signaled by
 the emergence of localized structures (discrete breathers~\cite{flach08})  which act as bottlenecks and make the system a bad heat conductor.  
 In this unusual situation, nonequilibrium negative temperature states are characterized by strong energy fluctuations and
 occur when large imbalances are imposed at the boundary of the chain.
 
 On a more general ground,  several fundamental
questions concerning heat transport at negative temperatures still need to be clarified.
For example: can diffusive transport at negative temperatures be described by a Fourier law for sufficiently
small thermal imbalances? What are the properties of the related stationary states?
Here we  shed some light on these issues   by studying both deterministic and stochastic systems driven out of equilibrium by possibly negative-temperature imbalances. 
 
 We start from one of the simplest systems satisfying our requirements, i.e. a paramagnetic chain of spins evolving according to stochastic local updates that conserve the total energy. Due to the absence of interaction terms
in the Hamiltonian, analytical expressions for the stationary temperature profiles and for the thermal conductivity can
be derived. We also show that the passage from the infinite temperature point does not pose any intrinsic limit to
the Fourier problem. Indeed we show that stationary states where the inverse temperature profile changes its sign are 
consistently described.
We then move to the study of a deterministic Hamiltonian system, namely the case of a chain of coupled rotators with bounded kinetic energy~\cite{cerino15, baldovin17, baldovin18}. In this case, we show that transport is diffusive and that the regime of  negative-temperature Fourier transport is accessible as well. Finally, we consider a slightly more complex situation where negative-temperature steady states are obtained in conditions
of coupled transport, i.e. when two independent currents are steadily sustained in the system as in thermoelectricity. Specifically, we 
analyze the case of a mixed deterministic-stochastic Schr\"odinger equation, a model recently introduced as a particularly simple
example where (diffusive) coupled transport arises in a harmonic chain of coupled complex oscillators subject to conservative noise~\cite{iubini19}. 

To investigate this peculiar transport setup, we need to introduce the interaction of
the system with suitable boundary reservoirs that can sustain negative temperatures. As usual, a reservoir is a very
large (ideally infinite) system that can either release or absorb energy without modifying its thermodynamical state. 
For the spin system we introduce a simple implementation of stochastic reservoir, while for deterministic systems we make use
of generalized Langevin equations as discussed in~\cite{baldovin18}.

The paper is organized as follows. In Sec.~\ref{sec:fouriernegative} we review some general properties of the 
nonequilibrium setup in one-dimensional systems and discuss a formulation of the Fourier law  that is 
naturally 	extendable to the case of negative-temperature heat conduction.
In Sec.~\ref{sec:spin} we analyze the case of heat conduction in a spin chain, while  Sec.s~\ref{sec:ham}
and~\ref{sec:schr} we discuss the cases of the Hamiltonian rotator chain and the linear-stochastic Schr\"odinger
equation, respectively. Finally, Sec.~\ref{sec:concl} is devoted to conclusions and to a brief summary of open 
problems.

\section{Fourier law and temperature profiles at negative temperature}
\label{sec:fouriernegative}

The study of one-dimensional models whose boundaries are kept at some fixed temperatures $T_L$ and $T_R$, with $T_R \ne T_L$, is far from being a trivial task, and it is usually difficult to provide a satisfactory theoretical explaination for the observed phenomenology~\cite{lepri03,dhar08,lepri16}. A relatively simple situation is represented by systems which are close to equilibrium, i.e. whose boundary temperatures are almost equal: in this case it is reasonable to expect that, after a suitable thermalization time, the system will be found in a stationary state in which a local-equilibrium condition is satisfied, and the average energy flux $j$ does not depend on the position $x$. If an expicit relation between $j$, the temperature and its gradient is known, from the condition
\begin{equation}
 j(T(x),\partial_x T(x)) = const\,,
\end{equation} 
it is clearly possible to get an equation for the temperature profile. Such an explicit relation is usually not known, but one can still rely on the empirical Fourier law
\begin{equation}
\label{eq:fourier}
 j= -\kappa(T) \partial_x T\,,
\end{equation} 
where $\kappa(T)$ is the thermal conductivity; if $T_R$ and $T_L$ are close enough, $\kappa(T)$ can be considered constant along the system, so that a linear temperature profile can be expected. It is worth noticing that the same reasoning can be repeated in terms of the inverse temperature $\beta=1/k_B T$ (in the following we put the Boltzmann constant $k_B$ equal to $1$). From Eq.~\eqref{eq:fourier} one gets
\begin{equation}
\label{eq:fourierbeta}
 j = \frac{\kappa(1/\beta)}{\beta^2} \partial_x \beta = \gamma(\beta) \partial_x \beta\,,
\end{equation} 
where $\gamma=\kappa/\beta^2$ is the associated thermal transport coefficient: if the relative difference between $T_R$ and $T_L$ is small, also $\gamma(\beta(x))$ is almost constant along the chain, and the above equation insures that the $\beta$ profile, not surprisingly, is almost linear as well.

In the following, we will argue that Eq.~(\ref{eq:fourierbeta}) provides a consistent description of Fourier transport
even in the regime of negative temperatures. In this respect we recall that,
as already stated in 
the seminal paper by Ramsey~\cite{ramsey56}, the introduction of negative temperatures does 
not affect the basic principle ``heat flows from hot to cold'', provided that 
the temperature scale is expressed in terms of $\beta$ instead of $T$. As a 
consequence, one might reasonably expect that the qualitative behaviour of these 
systems is still described by Eq.~\eqref{eq:fourierbeta} when they are driven 
out of equilibrium from thermal baths  at the boundaries with different 
temperatures. If this is true, linear $\beta$ profiles should be expected not 
only when $\beta_L=1/T_L$ and $\beta_R=1/T_R$ are both positive or both negative 
(a result which could have been deduced from Eq.~\eqref{eq:fourier} as well), 
but also when one of the two baths is kept at positive temperature and the other one at 
negative temperature. Since this point might appear not obvious in the light of some criticisms
moved against the whole concept of negative temperatures~\cite{romero13, dunkel14},
it is worth mentioning a simple argument which suggests the linearity of $\beta$ profiles
in conditions close to equilibrium, without making explicit use of the Fourier law.

Let us consider a homogeneous lattice of interacting particles, whose length $L$ 
along the $x$ axis is much larger than the size of any transversal section, so 
that it can be considered as a one-dimensional system. Particles are not allowed 
to leave their sites, but they can interact via short-range forces, which allow 
for heat transfer within the bar. The system is completely isolated, except for the 
left and right ends, which are subject to the action of external thermal bath,
which keep them at inverse temperatures $\beta_L$ and 
$\beta_R$, respectively. Stationary 
out-of-equilibrium conditions are realized as soon as $\beta_R \ne \beta_L$.  
Assuming that the total number of sites is very large, we can divide the lattice into 
$N\gg1$ identical cells of linear size $\Lambda=L/N$, identified by integer indexes 
$\indice=1,...,N\gg1$, each of them still containing a large number of particles. The 
cell size $\Lambda$ must be much larger than the typical interaction radius of 
the particles and the transversal size of the bar. We expect each cell to be 
found in a state of local equilibrium, described by local thermodynamic 
variables as the energy $E_\indice$ and the entropy $S(E_\indice)$, regarded as a function 
of the energy of the cell. Notice that $S$ and $E_\indice$ scale with $\Lambda$. Let 
us consider a pair of consecutive cells $\indice$ and $\indice+1$: their energies $E_\indice$ and 
$E_{\indice+1}$ are very close but still different. We want to establish a ``degree of 
local nonequilibrium'' $G_\indice$ for the considered pair of cells. First we compute 
the \textit{a priori} probability $P_\indice$ of a configuration with the observed 
energy values $E_\indice$ and $E_{\indice+1}$, given the total energy $2E^{eq}=E_\indice+E_{\indice+1}$. 
Recalling Boltzmann formula
\begin{equation}
 S(E)= \ln W(E)\,
\end{equation} 
where $W(E)$ is the number of states accessible to the system at energy $E$, we get
\begin{equation}
 P_\indice= \exp[S(E_\indice)+S(E_{\indice+1})-2S(E^{eq})]\,.
\end{equation} 
Then we define
\begin{equation}
\label{eq:gi}
 G_\indice=-\log(P_\indice)=2S(E^{eq})-S(E_\indice)-S(E_{\indice+1})\,,
\end{equation} 
so that $G_\indice$ vanishes when the system is at equilibrium and it increases when
the equilibrium probability of the thermodynamic state decreases. By defining \begin{equation}
 \Delta_\indice=E_\indice-E^{eq}=E^{eq}-E_{\indice+1}
\end{equation} 
we can expand Eq.~\eqref{eq:gi} assuming that $\Delta_\indice$ is small, so to obtain
\begin{equation}
 G_\indice \simeq -\frac{\partial^2 S}{\partial E_\indice^2}\Delta_\indice^2\,,
\end{equation}
and we can introduce a corresponding global observable for the whole system as
\begin{equation}
 \mathcal{G}=\sum_{\indice=1}^{N} G_\indice\,.
\end{equation} 
Taking the continuum limit
\begin{equation}
  n \Lambda \to x\quad\quad\frac{E_n}{\Lambda} \to \varepsilon(x)\quad\quad\frac{\partial^2 S}{\partial E^2_n}\Lambda \to \frac{d \beta}{d \varepsilon}\Big|_{\varepsilon(x)}\quad\quad \frac{\Delta}{\Lambda}\to \frac{1}{2}\varepsilon'(x)\,,
\end{equation}
we get
\begin{equation}
\label{eq:gcal}
 \mathcal{G}\to\frac{1}{4}\int_{0}^L\,\left[\varepsilon'(x)\right]^2 \frac{d \beta}{d \varepsilon}\Big|_{\varepsilon(x)} \,dx\,.
\end{equation} 
In the above we have exploited the definition of inverse temperature
\begin{equation}
 \beta\left( E_n/\Lambda \right)=\frac{\partial S}{\partial E_n}\,,
\end{equation} 
where the derivative in the r.h.s. is taken at fixed number of particles.
It is a reasonable physical assumption that the stationary temperature profile 
of the one-dimensional system will minimize $\mathcal{G}$. Indeed, larger values 
of $\mathcal{G}$ correspond to less probable configurations. Such minimization 
can be realized with a variational approach, leading to the Euler-Lagrange 
equation
\begin{equation}
\label{eq:ode}
 \frac{d^2\beta}{d \varepsilon^2}\left[  \varepsilon'(x)\right]^2 + 2 \frac{d \beta}{d \varepsilon}\varepsilon''(x)=0\,.
\end{equation}
Once the function $\beta(\varepsilon)$ is known, Eq.~\eqref{eq:ode} is a 
(possibly difficult) second order ordinary  differential equation, which only 
needs to be completed with the conditions $\beta(\varepsilon(0))=\beta_L$ and 
$\beta(\varepsilon(L))=\beta_R$ to be solvable.

Let us notice that the system is close to equilibrium, so that $\varepsilon$ 
is not expected to vary too much along the lattice and we can exploit the 
linearization
\begin{equation}
\label{eq:betalin}
 \beta(\varepsilon)=\beta_0+c \varepsilon\,,
\end{equation} 
Eq.~\eqref{eq:ode} reduces to the condition $\varepsilon''(x)=0$, so that the 
energy profile is linear and so is $\beta(\varepsilon(x))$ (by virtue of 
condition~\eqref{eq:betalin}).
This result does not depend on the specific signs of the boundary temperatures.
 In the specific case where  $\beta_L$ and $\beta_R$
have the same sign, we can simply recover the
 linearity of the profile of $T$. Indeed
\begin{equation}
\label{eq:T}
 \frac{d^2T}{dx^2}=\frac{2}{\beta^3}\left( \frac{d \beta}{d x}\right)^2 - 
\frac{1}{\beta^2}\frac{d ^2\beta}{d x^2} \approx \frac{2}{\beta^3}\left( \frac{d 
\beta}{d x}\right)^2
\end{equation} 
is a quantity of order $(\beta_L-\beta_R)^2$; when 
the difference between the inverse temperatures of the baths is small, $d^2T/dx^2$ is negligible and 
the temperature profile is linear, as it would be expected from the empirical Fourier law~\eqref{eq:fourier}.
 As expected, this result is not valid  anymore when $\beta(x)$ in the r.h.s. of Eq.~(\ref{eq:T}) vanishes for some $x$,
a condition which can be realized when $\beta_L$ and $\beta_R$ have opposite signs.

As a final remark, we point out  that  the functional  $\mathcal{G}$ in Eq.~\eqref{eq:gcal}
is related to the form of the entropy production 
\begin{equation}
W= \int_0^Lj  \varepsilon'(x) \frac{d \beta}{d \varepsilon}\Big|_{\varepsilon(x)}dx\,,
\end{equation}
once a proportionality relation is established between the heat flux $j$ and the 
energy density gradient $\varepsilon'(x)$. Accordingly, in this regime, the minimization of $\mathcal{G}$ corresponds to 
the well known principle of minimum entropy production~\cite{klein54}.

\section{Spin chain}
\label{sec:spin}
The first experiments involving equilibrium states at negative 
temperature were realized by Purcell, Pound and Ramsey, who were studying systems of nuclear 
spins exposed to intense external field~\cite{purcell51, ramsey56}. In this regime 
interactions are negligible, so that the energy of the system can be written as 
\begin{equation}
\label{eq:spinenergy}
 \mathcal{H}(\vect{\sigma})=-\hh\sum_{\indice=1}^{N}\sigma_\indice
\end{equation}
where $\sigma_\indice=\pm1$ is the $\indice$-th spin (with $\indice=1,...,N$, $N\gg1$) and $\hh$ 
represents the external magnetic field, which is supposed to be homogeneous. If 
there is a large time-scale separation between the typical thermalization times 
of the internal dynamics and those needed for a complete equilibration with the 
environment, and we focus on intermediate time-scales, the system can be 
considered both isolated and at equilibrium~\cite{purcell51, pound51}. Due to energy conservation, 
only simultaneous flipping of pairs of spins with opposite signs are allowed~\cite{abragam58}.

Let us denote the number of positive and negative spins by $N_+$ and 
$N_-=N-N_+$, respectively: in the simple model~\eqref{eq:spinenergy} 
conservation of energy is equivalent to keeping $N_+$ (or $N_-$) fixed, so that 
we can conveniently write an expression for the global entropy as 
\begin{equation}
\begin{aligned}
 & S(N_+)=\ln\left( \frac{N!}{N_+! N_-!}\right)\\
 & \approx - N_+ \ln \left(\frac{N_+}{N}\right)+N_- \ln \left(\frac{N_-}{N}\right) + O(\ln N)
\end{aligned}
\end{equation} 
where we have exploited Stirling's approximation; the entropy per particle reads therefore
\begin{equation}
 s(p)=-p\ln(p) -(1-p) \ln \left( 1-p \right)
\end{equation} 
where $p=N_+/N$ is the density of positive spins.
Now we can take advantage of the invertible relation between the specific energy $\varepsilon$ and $p$, i.e.
\begin{equation}
\label{eq:evsn}
 \varepsilon=\hh(1-2p)\,,
\end{equation} 
which follows from Eq.~\eqref{eq:spinenergy}; we can thus write an explicit expression for $s(\varepsilon,\hh)$:
\begin{equation}
 s(\varepsilon,\hh)= -\frac{\hh- \varepsilon}{2\hh} \ln \left( \frac{\hh- \varepsilon}{2\hh}\right) - \frac{\hh+ \varepsilon}{2\hh} \ln \left(\frac{\hh+ \varepsilon}{2\hh}\right) \,.
\end{equation} 
Taking the derivative of the above equation with respect to $\varepsilon$, for a fixed value of the external field, we get the inverse (microcanonical) temperature:
\begin{equation}
\label{eq:betaspin}
 \beta(\varepsilon,\hh)=\frac{\partial s}{\partial \varepsilon} = \frac{1}{2\hh}\ln\left(\frac{\hh-\varepsilon}{\hh+\varepsilon}\right)\,.
\end{equation} 
The specific energy $\varepsilon$ ranges from $-\hh$ to $\hh$, corresponding to the 
extreme cases in which all spins are positive or negative, respectively. As 
a consequence, the value of $\beta$ ranges from $-\infty$ to $+\infty$. Negative 
temperatures correspond to positive-energy cases in which most spins are not 
aligned with the external magnetic field. Let us also notice that by 
substituting Eq.~\eqref{eq:evsn} into Eq.~\eqref{eq:betaspin} and inverting, one 
finds
\begin{equation}
\label{eq:posdensity}
 p=\frac{e^{\beta \hh}}{2 \cosh(\beta \hh)}\,,
\end{equation} 
which is consistent with the statistical interpretation of $p$ as the 
probability that a given spin is positive, once the inverse temperature $\beta$ 
of the system is known. Similarly, the magnetization density of the system, 
$m=(N_+-N_-)/N$, verifies
\begin{equation}
 \label{eq:magnetization}
m=2 p -1=\tanh(\beta \hh)\,.
\end{equation} 

So far we have just recalled known equilibrium results for systems of Ising spins subjected to a strong
external field. In the following we will consider a linear geometry, i.e. we will assume that 
the spins are placed on the sites of a one-dimensional lattice. Taking into 
account local conservation of energy, we impose that only pairs of (opposite) 
spins located on \textit{neighbour} sites can exchange energy by simultaneous 
flipping. In particular, we model the dynamics assuming that at regular time 
intervals one of the $2(N-1)$ \textit{ordered} pairs of neighbour spins is 
randomly chosen (i.e., the couples $(\indice,\indice+1)$ and $(\indice+1,\indice)$ are regarded as 
different choices): if the first spin of the pair is positive and the second is 
negative, they both flip; otherwise, nothing happens. We will also consider 
thermal baths on the boundary sites of the chain, characterized by inverse 
temperatures $\beta_L$ and $\beta_R$. If $\beta_R\ne\beta_L$ the system is out 
of equilibrium, and we can ask what is the shape of the temperature profile in 
this case. Notice that, in principle, one or both thermal baths can be characterized by 
negative temperature.

 If the system reaches a stationary state, local conservation of energy implies 
that the average energy flux must be constant along the lattice. The dynamics 
can be conveniently mapped into an exclusion process~\cite{iubini14}, in which all 
positive spins are replaced by particles and all negative spins by empty sites; 
particles can only move to a neighbour site if this is empty, and each particle 
carries an amount of energy equal to $-2\hh$. With this scheme in mind, the 
average energy flux between the $\indice$-th and  $(\indice+1)$-th site is given by 
\begin{equation}
\label{eq:jspin}
\langle j_\indice \rangle=-2\hh\left(p_\indice w_{\indice\to \indice+1} - p_{\indice+1} w_{\indice+1\to \indice}\right)\,,
 \end{equation} 
 where $p_\indice$ is the probability of finding a particle in the $\indice$-th site (or, 
equivalently, the average occupation number of the $\indice$-th site), while $w_{\indice\to 
k}$ is the transition rate from site $\indice$ to site $k$. The latter can be written 
in terms of the occupation probability as
\begin{equation}
\label{eq:wspin}
  w_{\indice\to k}=\frac{1}{\tau}\, \text{Prob}(\,k \text{ empty } |\, \indice \text{ occupied } ) \approx \frac{1-p_k}{\tau}\,,
 \end{equation} 
where $\tau$ is the average time interval between two consecutive extractions of 
the ordered pair $(\indice,k)$. Indeed, in our model, once the pair has been 
extracted, the transition only happens if site $k$ is empty. In the above 
equation we have assumed that the probability of finding a particle in the $k$-th 
site does not depend on the occupation of its neighbour site $\indice$: this simplifying 
hypothesis, somehow resembling Boltzmann's ``molecular chaos'' assumption, is 
actually not true, since the dynamics may induce correlations between the 
occupation numbers of neighbour sites; however we can reasonably expect such 
correlations to be negligible in typical conditions. The condition that $\langle j_\indice \rangle$ 
assumes the same value for each $\indice$ in the stationary state (i.e. $\langle j_\indice \rangle \equiv j$ $\forall \indice$) leads then to
\begin{equation}
 p_\indice (1-p_{\indice+1})- p_{\indice+1} (1-p_{\indice})=p_\indice-p_{\indice+1}=const.
\end{equation} 
  As a consequence, the  nonequilibrium occupation probability of the $\indice$-th site reads
  \begin{equation}
  \label{eq:profilep}
   p_\indice=p_L+\frac{\indice-1}{N-1}(p_R-p_L)
  \end{equation}
  where $p_L$ and $p_R$ are the average occupation numbers of the leftmost 
($\indice=1$) and rightmost ($\indice=N$) sites, fixed by the boundary conditions. Assuming 
local equilibrium, we can expect $p_\indice$ to be well described by 
Eq.~\eqref{eq:posdensity} for the density of positive spins, where $\beta$ has 
now the meaning of a local inverse temperature. The relation can be inverted 
numerically so to find the $\beta$ profile along the lattice.
  
 \begin{figure}
 \centering
\includegraphics[width=.6\columnwidth]{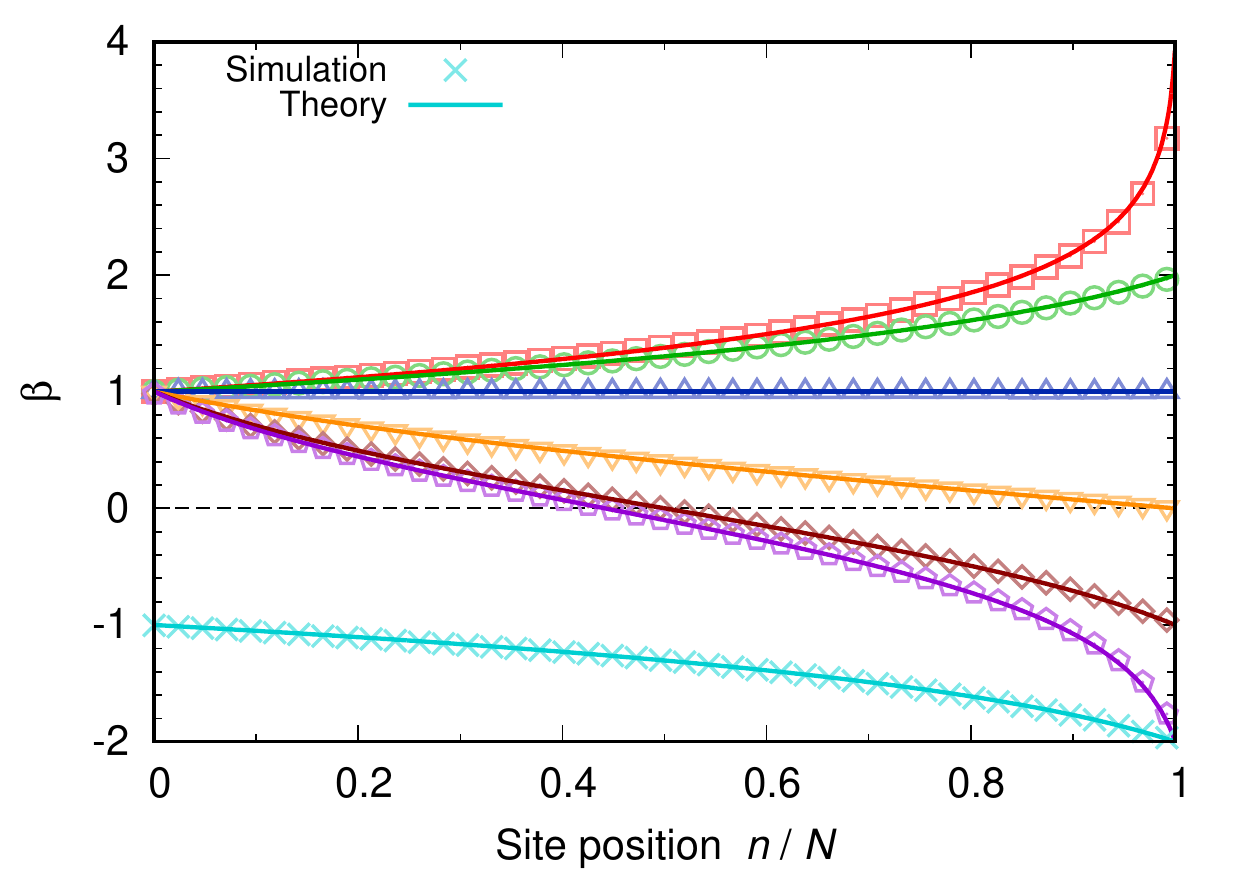}
\caption{\label{fig:spin2} Inverse temperature profiles for the spin chain, for different choices of the boundary conditions.
Points are obtained by measuring the local inverse temperature, solid lines are computed assuming that
Eq.~\eqref{eq:profilep} holds. Parameters: $\tau'/\tau=1$, $h=1$, $N=128$.}
 \end{figure}

  We can compare the above discussed results with the outcomes of numerical 
simulations in out-of-equilibrium conditions. The action of the thermal baths at 
the boundaries is mimicked by imposing that the probabilities $p_L$ and $p_R$
of a positive value of the leftmosrt and rightmost spin, respectively, are given by 
\begin{equation}
   p_{L,R}=\frac{e^{\beta_{L,R} \hh}}{2 \cosh(\beta_{L,R} \hh)}\,.
  \end{equation} 
  In this simple case, to impose the above distribution it is sufficient to
  update from time to time the extremal spins, extracting their values according
  to the above probabilities.
 The time intervals between consecutive updates are determined by a Poisson 
process with characteristic time $ \tau'$. We can then determining the local 
inverse temperature by measuring the local magnetization density and inverting 
relation~\eqref{eq:magnetization}. 
It can be verified that, if $\tau'/\tau \le 1$, the external baths are able to keep
the temperatures of the extremal sites fixed. The temperature profiles
for various choices of $\beta_L$ and $\beta_R$ are shown in 
Figure~\ref{fig:spin2}, where the theoretical expectations are also plotted.
 
 Let us remark that the presence of negative values of $\beta_R$ and/or 
$\beta_L$ does not hinder the possibility to reach a stationary state, 
characterized by constant energy flux along the chain; this is a first, simple 
example of Fourier-like transport in negative-temperature conditions. We 
also notice that, consistently with the discussion is 
Section~\ref{sec:fouriernegative}, the inverse temperature profile is continuous 
when passing from positive to negative values; the temperature profile would 
show instead a singularity when $\beta=0$.

We can also measure the ``conductivity'' $\gamma$ introduced in Eq.~\eqref{eq:fourierbeta},
which, at variance with $\kappa$, is well defined even when the boundaries of the chain are found at
temperatures with different signs.  To this end, from Eqs.~\eqref{eq:jspin} and~\eqref{eq:wspin} 
we can derive an explicit expression for the average value of the ``total'' heat flux 
\begin{equation}
 J = \sum_{\indice=1}^N j_{\indice}\,,
\end{equation} 
namely
 \begin{equation}
 \langle J \rangle = \frac{2\hh}{\tau}(p_R-p_L)\,.
\end{equation}
By using Eq.~\eqref{eq:posdensity}  and expanding to the linear order in $\Delta \beta=\beta_R-\beta_L$, we obtain
\begin{equation}
 \langle J \rangle \simeq \frac{1}{\tau}\left(\frac{\hh}{\cosh(\beta \hh)}\right)^2 \Delta \beta +O(\Delta\beta^2)\,.
\end{equation}
Accordingly,
\begin{equation}
\label{eq:gamma}
\gamma = \lim_{\Delta\beta\rightarrow 0}\frac{J}{\Delta\beta}=\frac{1}{\tau}\left( \frac{\hh}{\cosh(\beta \hh)}\right)^2\,.
\end{equation} 
  
\begin{figure}
 \centering
\includegraphics[width=.6\columnwidth]{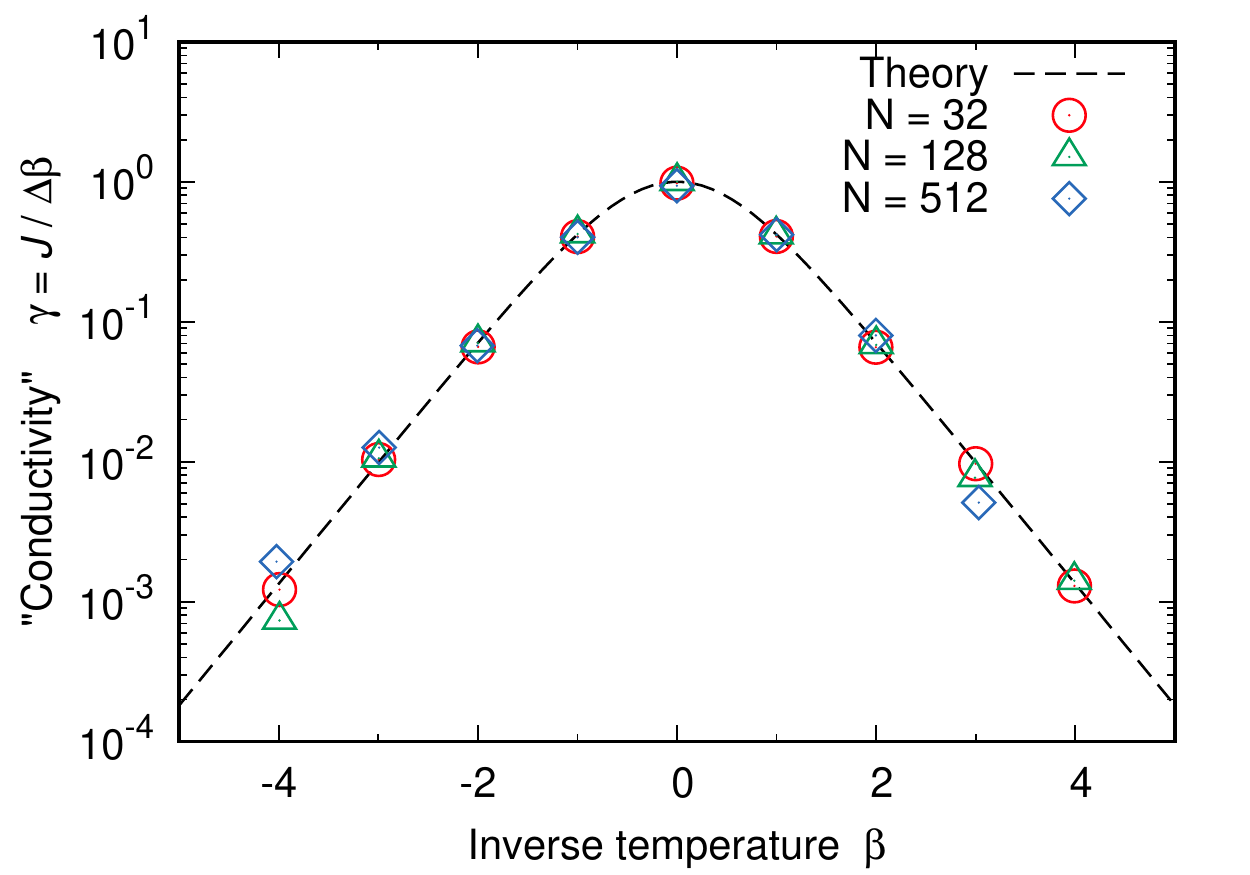}
\caption{\label{fig:spin3} Conducibility of the spin chain as a function of the inverse temperature $\beta$.
Points are obtained by measuring the energy flux along spin chains whose boundaries are kept at fixed inverse temperature $\beta \pm 0.05$, for three different values of $N$. The dashed line refers to the theoretical prediction~\eqref{eq:gamma}.
Parameters: $\tau'/\tau=1$, $\hh=1$.}
 \end{figure}

In Fig.~\ref{fig:spin3} we compare the measured values of $\gamma$ obtained from numerical simulations with the analytical expression Eq.~\eqref{eq:gamma}.
The reported plots show  a nice agreement for different values of the ``operating'' temperature
$\beta$, also in the negative-temperature region. 
Interestingly, the condition of maximal conductivity is obtained precisely at $\beta=0$, i.e. when the two thermal baths
operate at opposite inverse temperatures.

\section{Hamiltonian coupled rotators}
\label{sec:ham}

Negative temperature can be also found in Hamiltonian chains of the form
 \begin{equation}
\label{eq:hamchain}
  \mathcal{H}(\mathbf{p},\mathbf{q})=\sum_{\indice=1}^{N} \mathcal{K}(p_\indice) + \sum_{\indice=1}^{N-1} \mathcal{U}(q_{\indice+1}-q_\indice)\,.
 \end{equation}
 where the canonical variables $\{q_\indice,p_\indice\}$ live in a $2N$-dimensional bounded 
phase space, while $\mathcal{K}(p_\indice)$ represents a generalized kinetic term and 
$\mathcal{U}(q_{\indice+1}-q_\indice)$ is some interaction potential. In the following we 
will specialize to the case 
\begin{equation}
\label{eq:hamchain2}
 \mathcal{H}(\mathbf{p}, \mathbf{q})=\sum_{\indice=1}^{N} (1-\cos p_\indice) + \varepsilon \sum_{\indice=1}^{N-1}[1-\cos (q_{\indice+1}-q_{\indice})]\,,
\end{equation} 
i.e. a chain of $N$ particles characterized by kinetic terms which are 
periodic in the momenta, and potential interactions typical of systems of 
classical rotators~\cite{giardina00, gendelman00}. The phase space  $[0,2\pi)^{2N}$ of the coordinate 
variables is bounded, and this allows for the occurrence of negative temperature 
states. The  equilibrium properties of models of this kind have been 
investigated in previous works~\cite{cerino15, baldovin17, baldovin18}. Here we 
are interested in transport phenomena due to the presence of a temperature 
gradient. To this end, as in the case of the spin chain discussed previously, we 
need to impose different values $\beta_L$ and $\beta_R$ of the inverse 
temperature at the boundaries (particles labeled by $\indice=1$ and $\indice=N$), by means
of suitable stochastic heat baths. The most natural way to impose a equilibrium 
distribution on the $\indice$-th particle is to implement a dynamics ruled by the 
\textit{generalized} Klein-Kramers equation
\begin{equation}
\label{eq:genkk}
 \begin{pmatrix}
  \dot{q}_\indice\\
  \dot{p}_\indice
 \end{pmatrix}
 =
  \begin{pmatrix}
  \partial_{p_\indice} \mathcal{K}\\
  -\partial_{q_\indice} \mathcal{H}-D_\indice\beta_i\partial_{p_\indice} \mathcal{K} +\sqrt{2D_\indice}\xi(t)\,,
 \end{pmatrix}
\end{equation}
where $\xi(t)$ represents a Gaussian delta-correlated noise and $D$ is a 
parameter determining a typical frequency for the thermal bath. It can be shown 
that this stochastic differential equation reproduces the effect of an actual 
mechanical bath composed by a large number of particles with smaller 
inertia~\cite{baldovin18}. A similar equation also holds if the bath is composed 
by a set of Ising spins ruled by a Glauber dynamics and interacting with the 
considered particle~\cite{baldovin19}.

We 
implement molecular dynamics simulations using the algorithm discussed in 
Ref.~\cite{miceli19}; this integration scheme is quasi-symplectic, meaning that 
it exactly reduces to a symplectic algorithm (second-order Verlet) in the limit of vanishing 
noise~\cite{melchionna07}. The integration step $\Delta t$ is chosen by imposing 
that, in the zero-noise limit, the value of total energy is conserved, with 
relative fluctuations of order $\simeq 10^{-5}$. Determining the optimal value 
for $D$ is a slightly less trivial task, since it is known that the actual 
effect of Langevin-like baths in Hamiltonian one-dimensional chains is related 
to the characteristic time scale of the stochastic dynamics~\cite{lepri03}. 
This problem is discussed in Appendix~\ref{sec:appendixstoch} in some detail.

We start our analysis by focusing on heat transport at positive temperatures.
A consistency check for our setup can be done by measuring the thermal conductivity
 \begin{equation}
\label{eq:kappa}
 \kappa = \frac{Nj}{\beta_L^{-1}-\beta_R^{-1}}\,,
\end{equation} 
where $j$ is the local value of the heat flux, which can be identified with a suitable mechanical observable
 following the discussion in Ref.~\cite{lepri03}; first, we define the local 
energy as 
 \begin{equation}
  h_\indice=\mathcal{K}_\indice(p_\indice)+\frac{1}{2}\left[ \mathcal{U}_{\indice}(q_{\indice+1}-q_\indice)+\mathcal{U}_{\indice-1}(q_\indice-q_{\indice-1})\right]\,,
 \end{equation}
 and we consider its time derivative
\begin{equation}
\begin{aligned}
 \dot{h}_\indice&=\frac{1}{2}\mathcal{U}_{\indice}'(q_{\indice+1}-q_\indice)[\mathcal{K}_{\indice+1}'(p_{\indice+1})+\mathcal{K}_{\indice}'(p_{\indice})]+\\
 &-\frac{1}{2}\mathcal{U}_{\indice-1}'(q_{\indice}-q_{\indice-1})[\mathcal{K}_{\indice}'(p_{\indice})+\mathcal{K}_{\indice-1}'(p_{\indice-1})]\,.
 \end{aligned}
\end{equation}
The above expression has to be compared with the definition of heat flux in a 
linear lattice
\begin{equation}
 \dot{h}_\indice+j_{\indice}-j_{\indice-1}=0\,,
\end{equation}
where a unitary distance between neighbour sites on the lattice is assumed.
 We can thus identify the local heat flux as
\begin{equation}
\label{eq:heatflux}
 j_\indice=-\frac{1}{2}\mathcal{U}_\indice'(q_{\indice+1}-q_\indice)[\mathcal{K}_{\indice}'(p_{\indice}) + \mathcal{K}_{\indice+1}'(p_{\indice+1})]\,,
\end{equation} 
which in our case specializes into
\begin{equation}
\label{eq:jham}
 j_\indice=-\frac{\varepsilon}{2}\sin(q_{\indice+1}-q_\indice)[\sin p_{\indice} + \sin p_{\indice+1} ]\,.
\end{equation}
In the stationary state, $j=\langle 
j_\indice \rangle$ is expected to be independent of the site $\indice$. 

\begin{figure}
 \centering
\includegraphics[width=.6\columnwidth]{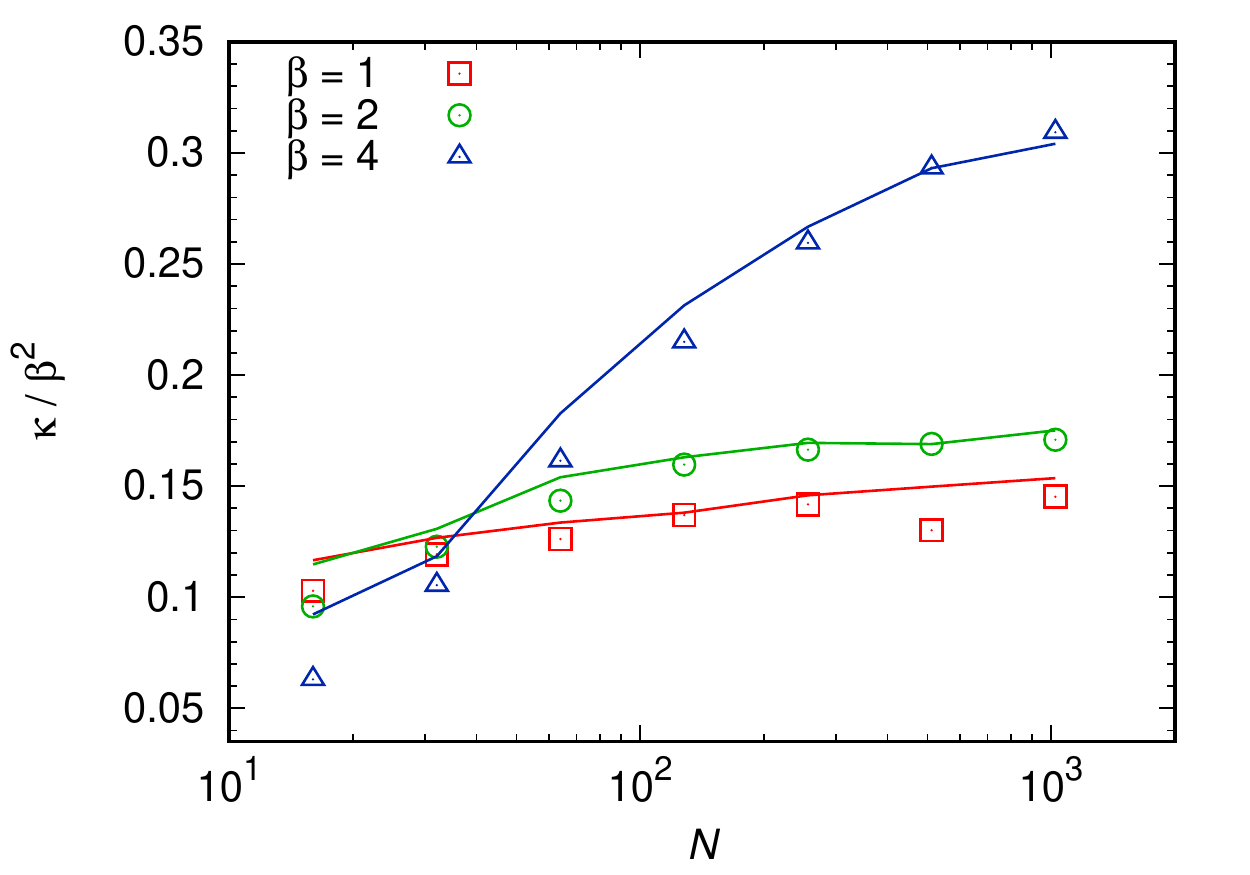}
\caption{\label{fig:ham3} Size-dependence of thermal conductivity for the rotators model~\eqref{eq:hamchain2}. For different values of $\beta$, the results of out-of-equilibrium numerical simulations in which $\beta_{L,R}=(1 \pm 0.05)\beta$ are shown. Points represent conductivity $\kappa$ as measured in numerical simulations from its definition~\eqref{eq:kappa}; solid lines are computed according to the right hand side of Eq.~\eqref{eq:kappagk} at finite $N$. Here $D=0.4$; other parameters as in Fig.~\ref{fig:ham1}.}
 \end{figure}

If the difference between the temperatures of the thermal baths is small, the total heat flux
\begin{equation}
\label{eq:totalj}
 J=\sum_{\indice=1}^{N}j_{\indice}
\end{equation} 
can be seen as the stationary response to a perturbation of the equilibrium 
distribution function, and a Green-Kubo relation between the conductivity and 
the autocorrelation of $J$ can be found~\cite{kubo66, lepri03}:
\begin{equation}
\label{eq:kappagk}
  \frac{\kappa_{GK}}{\beta^2} = \lim_{t \to \infty} \int_{0}^{t}\, d\tau\, \lim_{N \to \infty} \frac{\langle J(\tau) J(0)\rangle}{N}\,.
\end{equation} 
In Fig.~\ref{fig:ham3} we compare the values of $\kappa$ measured in numerical 
simulations to the prediction $\kappa_{GK}$ based on the Green-Kubo analysis, 
for fixed size of the chain (i.e., we do not take the $N\to \infty$ limit) in 
conditions close to thermal equilibrium. Let us notice, incidentally, that only 
for $\beta$ small enough the conductivity reaches a finite limit at large $N$ 
(normal conduction) -- or, at least, the value of $N$ at which $\kappa$ reaches 
a plateau seems to depend on the inverse temperature. This is consistent with 
known results for the case of classical rotators (characterized by the same 
potential terms as in Eq.~\eqref{eq:hamchain2}, but usual quadratic kinetic 
terms): in that case a transition between two regimes of transport is 
observed for $\beta \simeq 3$~\cite{gendelman00}.

\begin{figure}
 \centering
\includegraphics[width=.6\columnwidth]{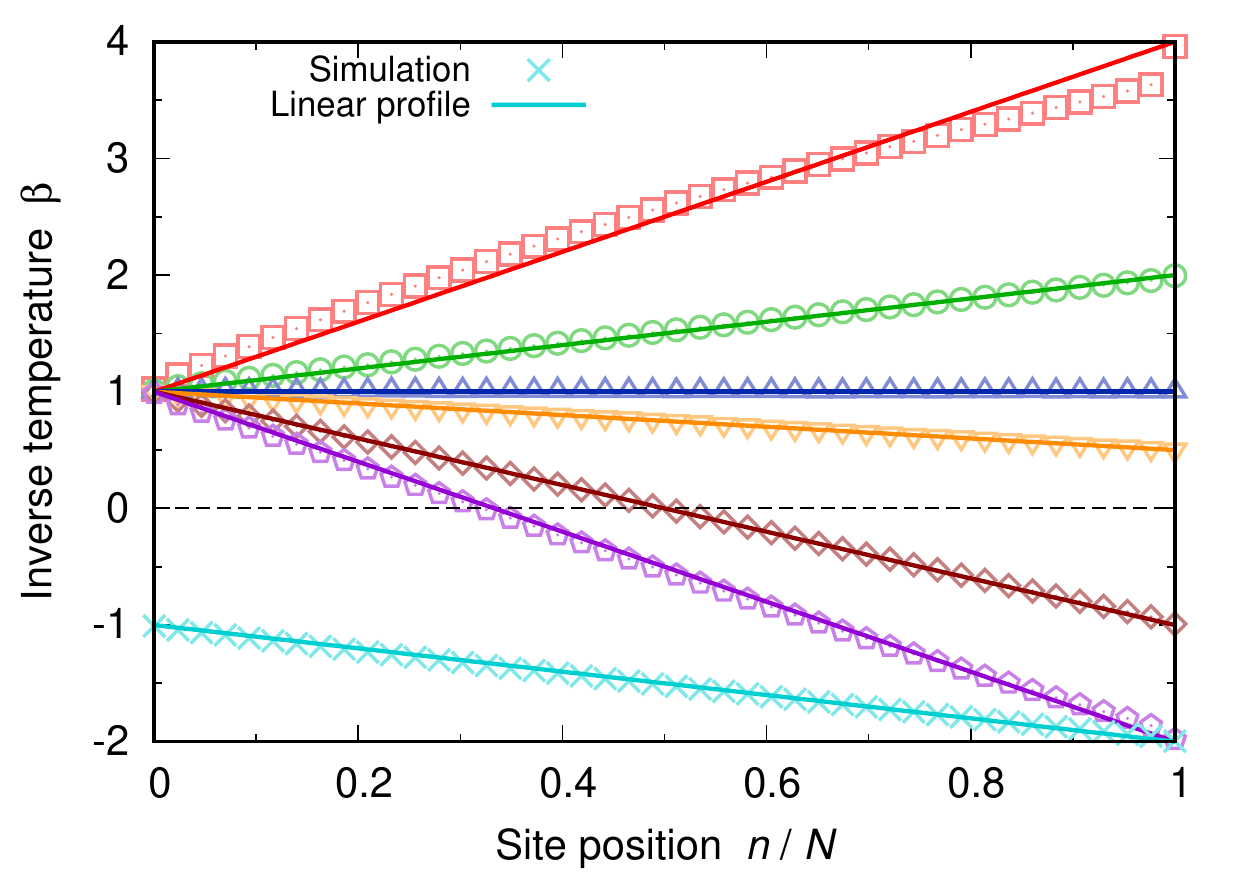}
\caption{\label{fig:ham4} Inverse temperature profiles for system~\eqref{eq:hamchain2} driven out of equilibrium. Points represent the outcomes of numerical simulations, solid lines the linear profile between the considered inverse temperatures. Parameters: $\varepsilon=0.5$, $N=1024$, $D=0.4$, $\Delta t=3 \cdot 10^{-2}$.}
 \end{figure}

 Finally, in Fig.~\ref{fig:ham4} we plot the local inverse temperature $\beta_\indice$ 
of the chain, as a function of the site $n$, for different values of $\beta_L$ 
and $\beta_R$. Here $\beta_\indice$ is measured by inverting the relation
 \begin{equation}
  \langle \cos p_\indice \rangle = \frac{ \int_0^{2\pi} d p_\indice\, e^{-\beta_\indice \mathcal{K}(p_\indice)}\, \cos p_\indice}{ \int_0^{2\pi} d p_\indice e^{-\beta_\indice \mathcal{K}(p_\indice)}} =\frac{I_1(\beta_\indice)}{I_0(\beta_\indice)}\,,
 \end{equation} 
 $I_n(x)$ being the modified Bessel function of the first kind. The inverse 
temperature profile does not significantly deviate from the linear behaviour 
predicted by the simple argument in Section~\ref{sec:fouriernegative}. Let us 
stress once again that this is also true when one or both temperatures are 
negative; as a consequence, the crossing between the positive- and 
negative-temperature regime occurs at $\beta=0$, suggesting once again that this 
state is physically meaningful and may be realized in out-of-equilibrium chains 
allowing for negative temperature states.

\section{Linear-Stochastic Schr\"{o}dinger Equation}
\label{sec:schr}

Fourier transport at negative temperatures may arise even in more complex setups as those involving coupled transport,
i.e. when two or more species of currents flow through the system and influence one another~\cite{livi17}. 
The study of coupled transport phenomena in chains of coupled oscillators is a relatively recent topic~\cite{iubini12,iubini14,borlenghi15,iubini16,wang20,iacobucci20}
In this section we will focus on a minimalist model of a chain of coupled Schr\"odinger oscillators defined by
the equations

\begin{equation}
\label{eq:dls}
i\dot z_\indice = -z_{\indice+1} - z_{\indice-1} \quad \indice=1,\cdots,N
\end{equation}
where the variables $z_n$ and $i z_\indice^*$ are a couple of complex-valued canonical variables for the Hamiltonian

\begin{equation}
H=\sum_{\indice=1}^N   z_{\indice}^* z_{\indice+1} + z_\indice z_{\indice+1}^*
\end{equation}

and open boundary conditions $z_0=z_{N+1}=0$ are assumed. This model can be regarded as the
linear limit of the celebrated Discrete Nonlinear Schr\"odinger equation~\cite{kevrekidis09}.
Given its linear structure, the model can be exactly diagonalized, giving rise to $(N-1)$ additional conserved quantities
besides energy. Among them, the conservation of the total norm

\begin{equation}
A = \sum_{n=1}^N |z_n|^2
\end{equation}

is related to the invariance of Eq.~(\ref{eq:dls}) under global phase transformations of the $z_\indice$ in the complex plane.
While the dynamics of Eq.~\ref{eq:dls} is completely integrable and gives rise naturally to ballistic transport, 
in ~\cite{iubini19} it was shown that in the presence of a suitable external conservative noise, the set of conserved
quantities can be reduced to the couple $(A,H)$. The irreversible dynamics amounts to the addition of conservative
``collisions'' among neighbouring oscillators occurring at rate $\nu_c$.
 Such collisions can account, for example, for the effect of a weak nonlinearity of the equations of motion~\cite{onorato20}.
 In detail, given a randomly selected site 
$\indice=2,\cdots,N-1$ in the bulk of the chain, a collision event consists in modifying the phase $\phi_\indice=\arg(z_n)$ such that 
the local energy density $h_n =\left(z_{\indice}^* z_{\indice+1} + z_\indice^* z_{\indice-1} +c.c.\right)/2 $ is conserved. In addition, this pure
phase transformation conserves the local norm $|z_n|^2$ and therefore $A$.
Collision events occur at random times, whose separations $\tau$ are independent and identically distributed random variables
extracted from a Poissonian distribution $P(\tau)\simeq \exp(-\nu_c \tau)$. Given its  mixed linear-deterministic and stochastic
evolution, we will refer to the model as the Linear-Stochastic Schr\"odinger (LSS) equation.  
 
\begin{figure}
 \centering
\includegraphics[width=0.6\columnwidth]{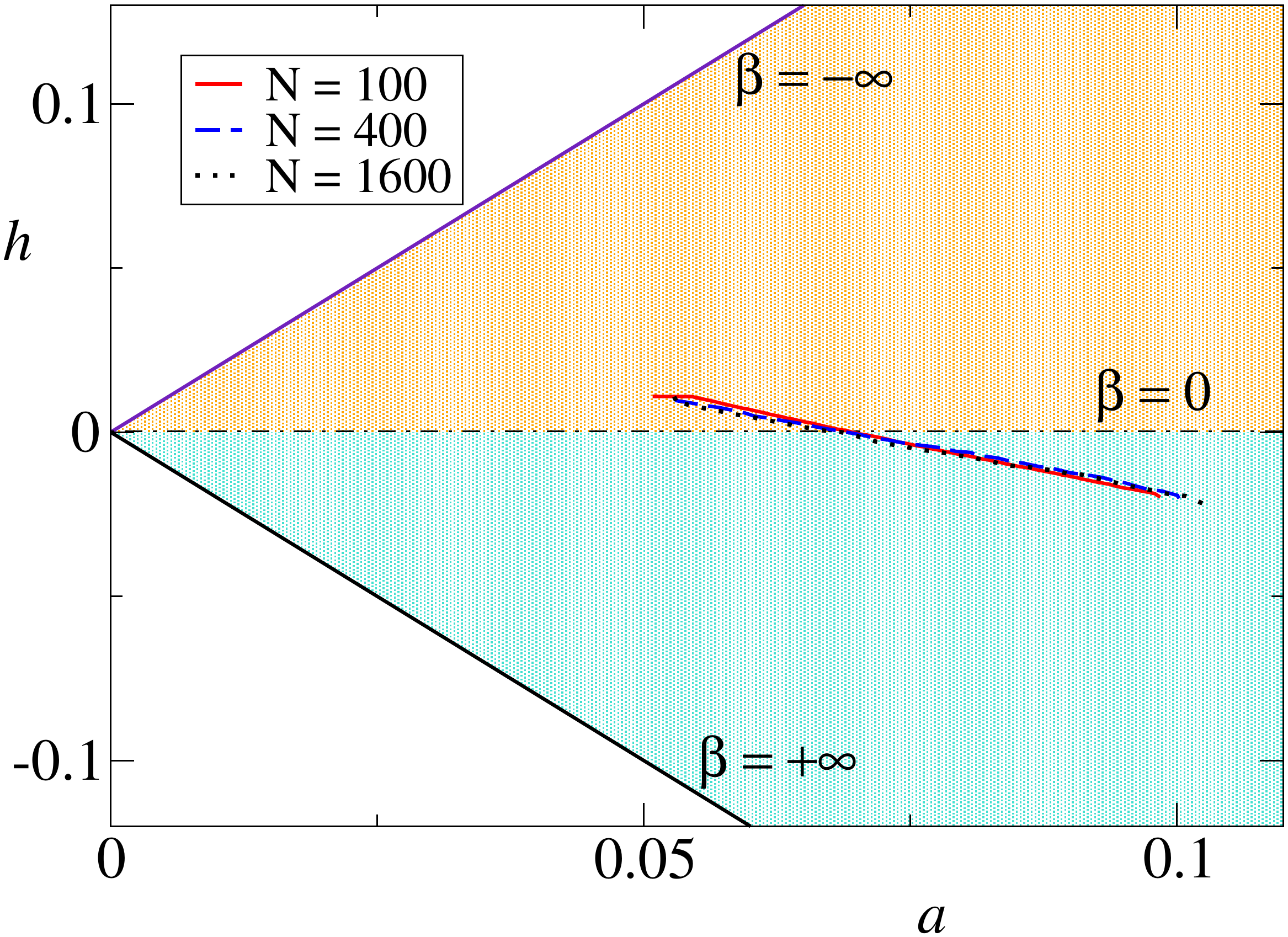}
\caption{\label{fig:DLS0} 
Thermodynamic diagram $(a,h)$ of the LSS equation. Black and purple solid lines are, respectively, the ground state ($\beta=+\infty$)
and the $\beta=-\infty$ state. The horizontal dot-dashed curve identifies the $\beta=0$ line. Parametric plots $[a(x),h(x)]$ of
nonequilibrium  density profiles are shown for $\beta_L=-2$, $\beta_R=1$, $\mu_L=10$ and $\mu_R=-10$. }
 \end{figure}

The equilibrium properties of the LLS model were discussed in
~\cite{buonsante16,iubini19} within the grand-canonical ensemble and they can be summarized in the phase diagram shown in Fig.~\ref{fig:DLS0}, which involves 
the norm density $a=A/N$ and the energy density $h=H/N$. 
The solid black and purple lines, defined by $h=\mp 2a$, represent
 respectively the ground  state $(\beta=+\infty)$ and the maximum energy state $(\beta=-\infty)$, while the horizontal dot-dashed line is the $\beta=0$ isothermal. 
 The region of parameters above this line is characterized by  negative absolute temperatures, see orange  area.
Equilibrium states at positive temperatures lie below the $\beta=0$ line, see turquoise region.

The transport properties of the model were studied in~\cite{iubini19} within the positive temperature region. In particular,
it was found that transport is diffusive and that the Seebeck coefficient is  non-vanishing, thus revealing the existence of coupled-transport~\footnote{In the context of linear response theory, the Seebeck coefficient $S$ corresponds to minus the ratio between the chemical potential gradient and the temperature gradient in the absence of norm flux, see~\cite{iubini12} for details. Coupled transport is identified by the condition $S\neq 0$.}.
  In the following we will show that the above model displays Fourier transport
also at negative temperatures.

The study of the nonequilibrium problem can be accomplished by
imposing suitable imbalances  of inverse temperature $\beta_R-\beta_L$ and of chemical 
potential $\mu_R-\mu_L$ at the boundaries of the chain.  
To this end, we will consider  the following Langevin equation specified for the 
left reservoir 	 connected to the lattice site $\indice=1$

\begin{equation}
\label{eq:dls_lang}
i\dot z_1 = -(1+i\alpha_L)z_2  +i\alpha_L\mu_L z_1 + \sqrt{\alpha_L / \beta_L}\, \eta(t)\,,
\end{equation}
where $\alpha_L$ specifies the coupling parameter and $\eta(t)$ is a (complex) Gaussian white noise with
zero mean and unit variance satisfying $\langle\eta(t)\eta(t') \rangle=\delta(t-t')$. An analogous equation
holds for the right reservoir acting on site $n=N$ and characterized by parameters $\alpha_R,\beta_R,\mu_R$.
The derivation of Eq.~(\ref{eq:dls_lang}) is discussed in~\cite{iubini13off} assuming that $\beta_L$ and $\alpha_L$
are both positive-definite. In analogy with~\cite{baldovin18} it can be formally extended to the case $\beta_L<0$ by
requiring  $\alpha_L<0$, thus replacing the standard dissipative term with a gain.

In addition to the boundary thermodynamic parameters, one needs to perform measurements of inverse temperature
and chemical potential profiles. This task is accomplished for the LSS equation by making use of suitable microcanonical observables
derived from the thermodynamic relations $\beta=\partial \mathcal{S} / \partial E|_{A=M}$ and $-\beta\mu=\partial \mathcal{S}/\partial A|_{H=E}$,
for a system with total energy $H=E$, total norm $A=M$ and entropy $\mathcal{S}$, see~\cite{franzosi11b,iubini12} for details.

\begin{figure}
 \centering
\includegraphics[width=0.6\columnwidth]{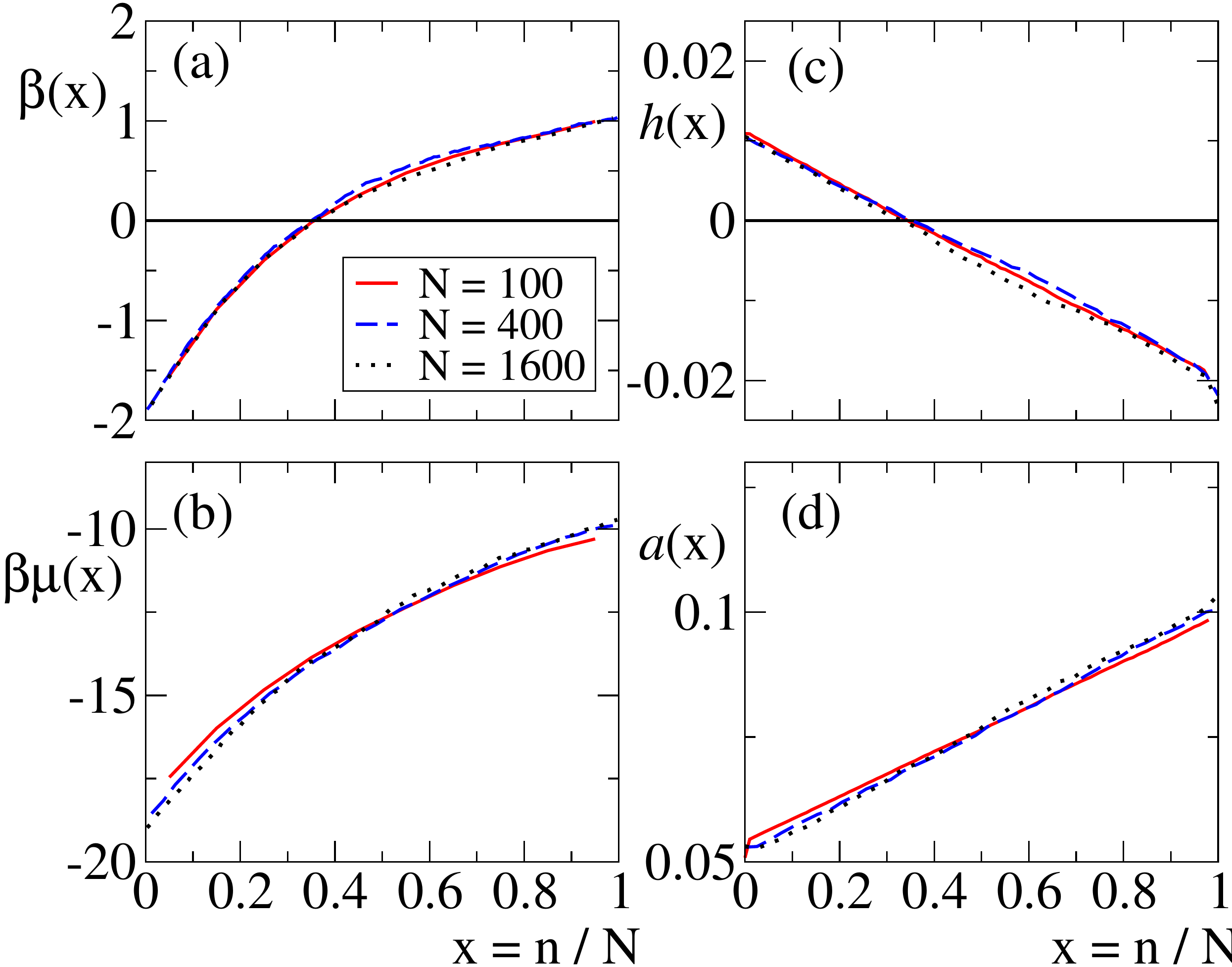}
\caption{\label{fig:DLS1} 
Nonequilibrium stationary profiles a  LSS chain of $N=100,400$ and $1600$ sites and parameters 
 $\beta_L=-2$, $\beta_R=1$, $\mu_L=10$ and $\mu_R=-10$. Simulations are obtained evolving the LSS system for
$5\cdot 10^6$ time units after a  transient of $10^6$ time units with parameters  $\nu_c=1$, $\alpha_L=-1$ and $\alpha_R=1$.}
 \end{figure}

In Fig.~\ref{fig:DLS1} we show  nonequilibrium stationary profiles corresponding to a LSS equation in contact with boundary 
reservoirs with inverse temperature $\beta_L=-2$ and $\beta_R=1$ with chemical potentials $\mu_L=10$ and $\mu_R=-10$.
We have chosen  $\nu_c=1$, $\alpha_L=-1$ and $\alpha_R=1$.
All profiles display an extensive scaling for different system sizes $N$ and they nicely overlap once reported as a function 
of the intensive spatial variable $x=\indice/N$.
As discussed for the previous models, the profiles of $\beta(x)$ cross smoothly the $\beta=0$ point (panel (a)).
The same is observed for the profiles of $(\beta\mu)(x)$ (panel (b)), where such a crossing occurs for a finite value of $\beta\mu$.
This is consistent with the fact  that on the $\beta=0$ line in Fig.~\ref{fig:DLS0} the chemical potential has a discontinuity $\mu=\pm \infty$. 
Therefore, it appears that the second appropriate thermodynamic variable  (besides $\beta$) to describe coupled transport 
across $\beta=0$ is  $\beta\mu$. The corresponding profiles of local  energy $h(x)$  and  norm $a(x)$ are shown in panels (c) and (d), 
respectively.  

The same density profiles  are represented parametrically in the diagram $(a,h)$ in Fig.~\ref{fig:DLS0}, where
they correspond to overlapping linear lines that connect the regions of positive and negative temperatures.

\begin{figure}
 \centering
\includegraphics[width=.6\columnwidth]{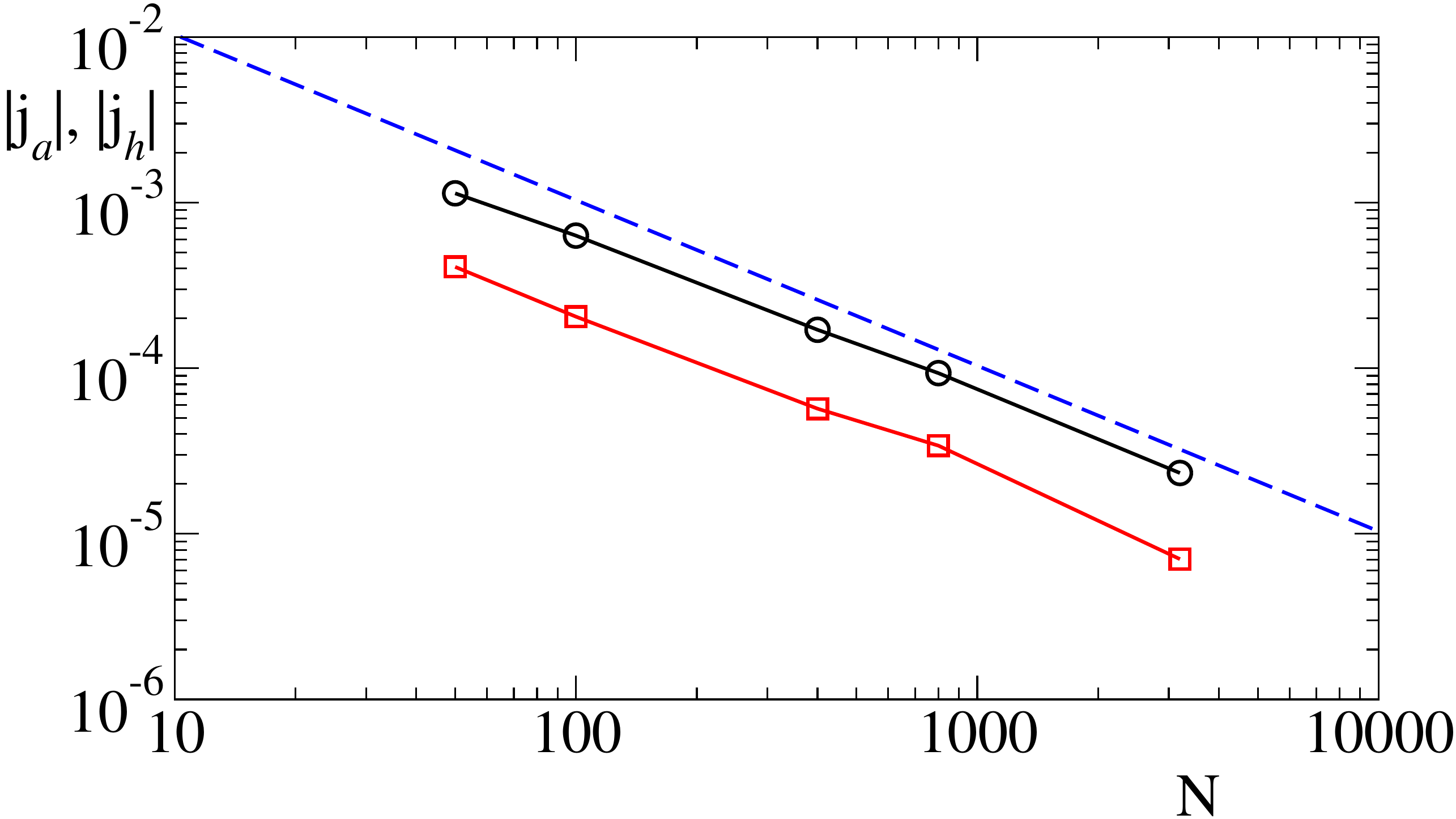}
\caption{\label{fig:DLS2} 
Average norm flux (black circles) and energy flux (red squares) versus chain size $N$  for a  LSS system with $\beta_L=-2$, $\beta_R=1$, $\mu_L=10$ and $\mu_R=-10$.
The blue dashed line shows the scaling  $1/N$.}
 \end{figure}

To better analyze the properties of the transport process, we have measured the stationary currents of norm and
energy, which are respectively defined~\cite{iubini12,iubini19} as
\begin{equation}
j_a=2\langle {\rm Im}(z_\indice^* z_{\indice+1})\rangle, \quad \,j_h=2\langle {\rm Re}(\dot z_\indice z_{\indice+1}^*)\rangle\,,
\end{equation}
where the symbol $\langle \cdot \rangle$ represents the average over the nonequilibrium stationary distribution.
These currents are reported (in absolute value) in Fig.~\ref{fig:DLS2} as a function of the system size $N$. A clear dependence as
$1/N$ is observed for both currents, thus confirming that Fourier coupled transport is established.

\section{Conclusions}
\label{sec:concl}

We have presented a study of  thermal transport in one-dimensional models  compatible with negative absolute temperatures and we have performed a first exploration of their stationary states in the negative temperature region.
We have considered three different types of microscopic dynamical evolution, namely a purely stochastic spin chain, a 
deterministic Hamiltonian model of rotators and a mixed deterministic-stochastic model of complex coupled oscillators. While
the first two models are characterized by transport of a single (heat) current, the third one admits two independent currents, hence coupled transport in the sense of irreversible thermodynamics. 
 
The necessity to impose given values of the thermodynamic parameters at the chain ends requires the introduction
of external reservoirs working both at positive and negative temperatures.  Such reservoirs have been implemented 
in the spin chain by simply imposing the target equilibrium distribution on the extremal spins,
and in the oscillators model as a Langevin equation complementing the deterministic dynamics. 

We have shown that the natural thermodynamic parameters for the description of negative-temperature transport are the inverse temperature $\beta$
and, when coupled transport is present, the quantity $\beta\mu$.
Within this representation, no intrinsic singularities occur upon crossing the infinite-temperature ($\beta=0$) point.
Indeed, our results show that stationary transport regimes connecting the positive- and negative-temperature regions are
physically accessible. In this generalized transport setup, a form of the Fourier law as in~\eqref{eq:fourierbeta} describes consistently diffusive transport through  the conductivity parameter
$\gamma$, see also Eq.~(\ref{eq:gamma}) in the whole space of accessible thermodynamic parameters.
 We remark that for the models here considered, $\gamma$ turns out to be finite in the  limit of large system sizes $N$, i.e.
 transport is normal. 

Among the possible perspectives, a natural one is to identify and study further models of negative-temperature
transport. This task is less obvious than one can imagine. Indeed, beyond the identification of a model where negative 
temperatures are accessible, one has to verify that the coupling with external reservoirs at negative temperatures 
is compatible with the physics of the  system. A relevant example where this program
breaks down is the Discrete Nonlinear Schr\"odinger equation~\cite{kevrekidis09}. Recent studies have shown that this system displays negative temperature states only 
within the microcanonical ensemble, while a grand-canonical description is forbidden because of nonequivalence of statistical
ensembles~\cite{GILM1,GILM2}. As a result, the system can not be consistently driven to the negative-temperature region by means of a negative-temperature reservoir. On the one hand, this does not exclude the occurrence of negative-temperature transport by 
 large imbalances of positive-temperature reservoirs, as shown in~\cite{iubini17entropy}. On the other hand, the joint observation of
strongly inhomogeneous profiles  appears to forbid any description of the process in terms of generalized conductivity coefficients.

\appendix

 \begin{figure}
 \centering
\includegraphics[width=.6\columnwidth]{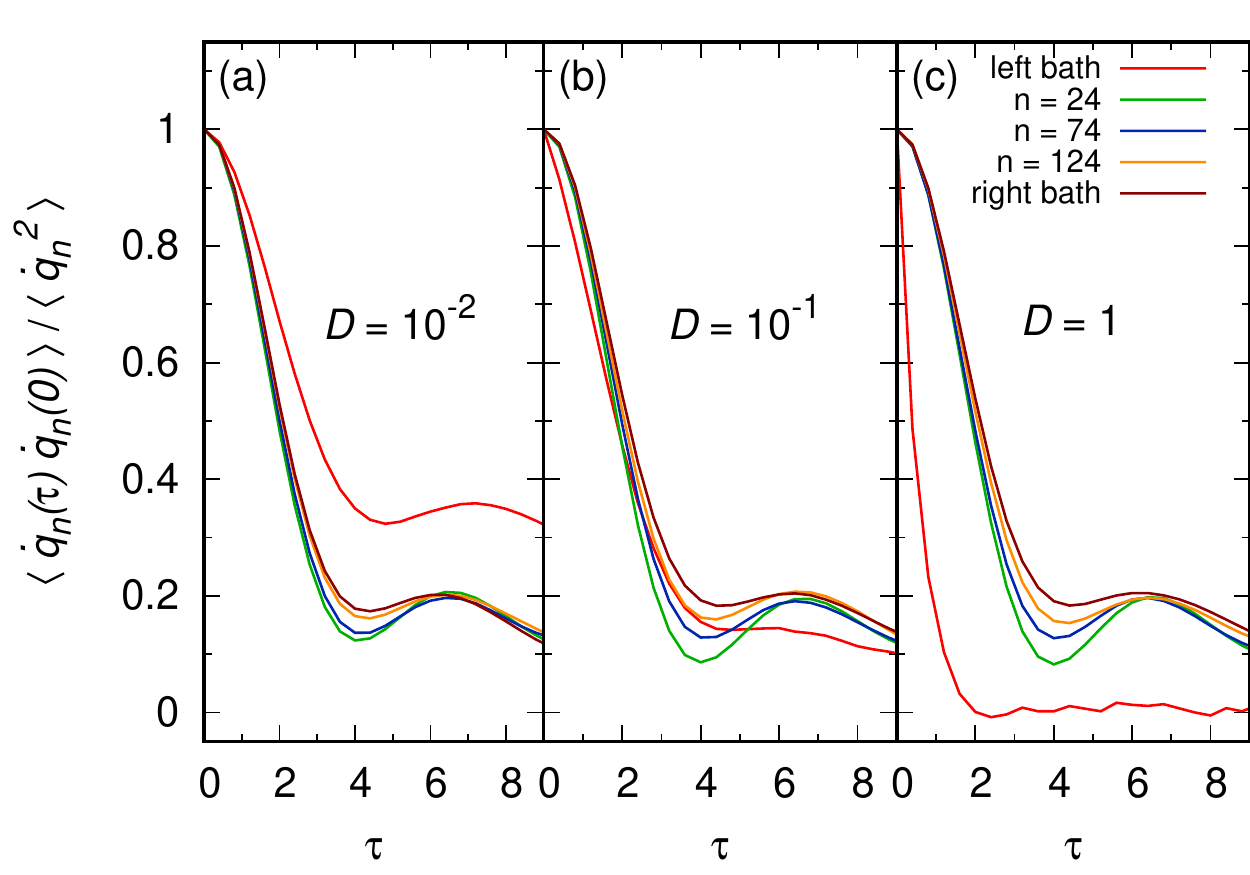}
\caption{\label{fig:ham1} Autocorrelation functions of the rotators model~\eqref{eq:hamchain2}  for some particle velocities. Cases $D=10^{-2}$, $D=10^{-1}$ and $D=1$ (panels (a), (b) and (c), respectively), are considered: only for the second choice the characteristic times of the boundary particles (subjected to the action of the baths) are comparable with those of the bulk. Parameters: $\varepsilon=0.5$, $N=128$, integration step $\Delta t =3 \cdot 10^{-2}$.}
 \end{figure}

 \begin{figure}
 \centering
\includegraphics[width=.6\columnwidth]{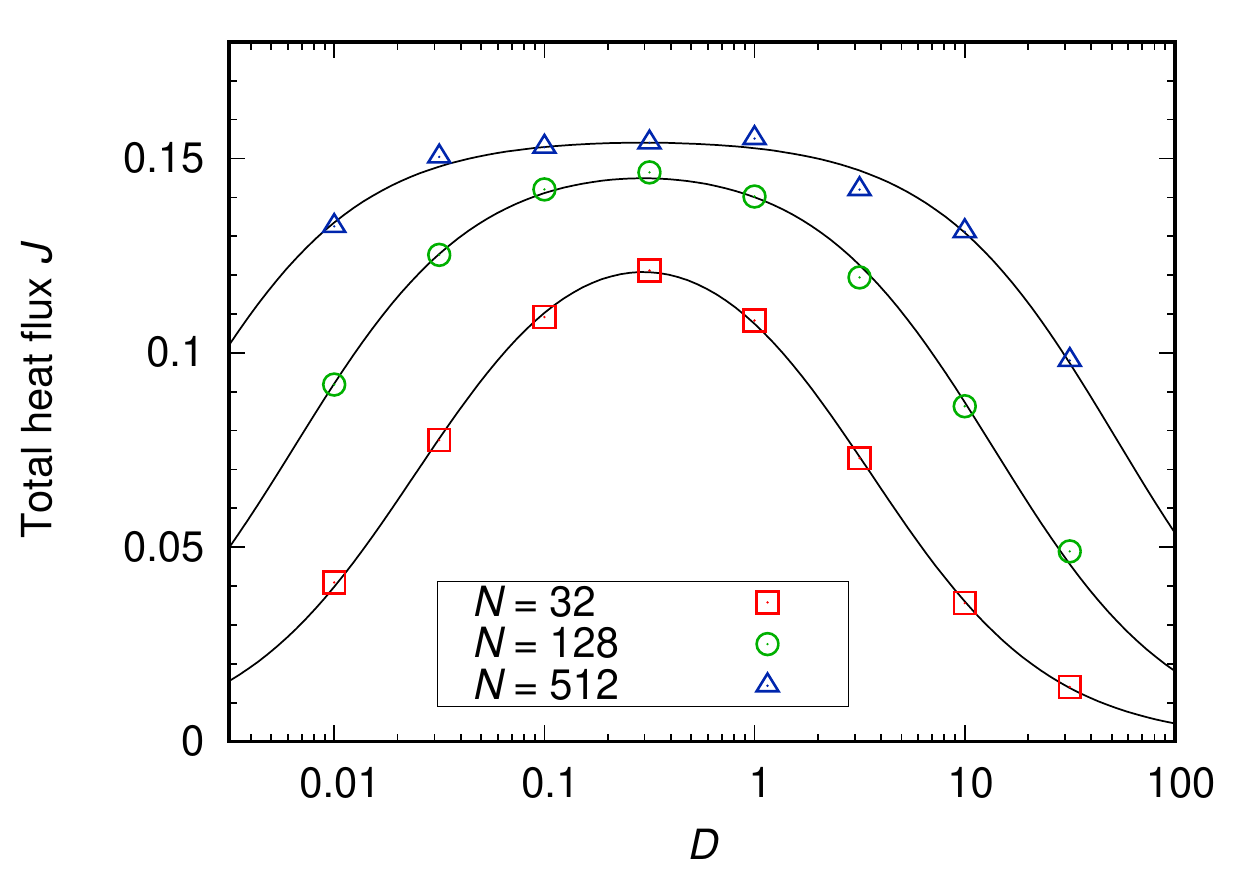}
\caption{\label{fig:ham2}Dependence on $D$ of the total heat flux  of the rotators  model~\eqref{eq:hamchain2}. For 3 values of $N$,
the behaviour of $J$ as a function of $D$ is shown (points). Black solid lines are obtained with a least-square
fit of the heuristic functional form~\eqref{eq:heuristic}. Here, $\beta_L=1$, $\beta_R=2$. Other parameters as in Fig.~\ref{fig:ham1}.}
 \end{figure}

\section{Remarks on Langevin-like stochastic baths}
\label{sec:appendixstoch} 

In this Appendix we study the problem of finding the optimal value for the 
paramter $D$ which characterize the thermal baths in Eq.~\eqref{eq:genkk}. When 
Langevin-like reservoirs are employed in numerical simulations, fixing the value 
of the parameters appearing in the stochastic differential equations amounts to 
determine a typical time scale for their dynamics. If one aims to understand
the equilibrium properties of a given system, this fact can be regarded as a 
secondary aspect, since equilibrium average are not expected to depend on such 
dynamical properties. The scenario is different when out-of-equilibrium 
conditions are considered, since one also needs that the chosen parameters
actually reproduce the right dynamics. 

 For the case of the Hamiltonian rotator model~\eqref{eq:hamchain2}, some insight on the best choice for $D$ can be gained by looking at 
the autocorrelation functions of some particles for different choices of the 
parameters, as shown in Fig.~\ref{fig:ham1}: for too large (or too small) values 
of $D$ the characteristic times of the bulk can be very different from those of 
the boundary particles, a clear hint that the stochastic dynamics does not 
reproduce the typical behaviour of the Hamiltonian particles in this case; 
better agreement is found for $D \simeq 10^{-1}$.

A more quantitative analysis 
can be drawn by studying the behaviour of the heat flux~\eqref{eq:totalj}. The 
behaviour of $J$ as a function of $D$ is shown in Fig.~\ref{fig:ham2} for some 
values of $N$; it is nicely fitted by the empirical formula
\begin{equation}
\label{eq:heuristic}
 J \approx \frac{c_1 D}{1+c_2D+c_3D^2}\,,
\end{equation} 
resembling the one discussed in~\cite{lepri03}. When $N$ large enough, a plateau 
is observed for values of $D$ between $0.1$ and $1$, which maximize the heat flux.
For our numerical simulations we chose values of $D$ in such interval.

\begin{acknowledgments}
We acknowledge useful discussions with Roberto Livi and Angelo Vulpiani. We thank Stefano Lepri 
for a critical reading of the manuscript.
M. B. acknowledges financial support of MIUR-PRIN2017 \textit{Coarse-grained description for non-equilibrium systems and transport phenomena (CO-NEST)}. 
\end{acknowledgments}

\bibliography{biblio}

\end{document}